\documentclass[twocolumn]{aastex631}
\usepackage{apjfonts}
\usepackage{graphicx}
\usepackage{amsmath}
\usepackage{amssymb}
\usepackage{color}

\gdef\xx[#1]{\textcolor{red}{#1}}

\gdef\kms{km\,s$^{-1}$}
\gdef\msun{M$_{\odot}$}

\gdef\lya{Ly\kern 0.09em$\alpha$}
\gdef\ha{H\kern 0.09em$\alpha$}

\newcommand{\GG}[1]{}

\newcommand{\RA}[3]{#1$^\mathrm{h}$#2$^\mathrm{m}$#3$^\mathrm{s}$}
\newcommand{\DEC}[4]{#1#2$^\circ$#3'#4''}

\lefthead{van Dokkum et al.}
\righthead{}
\begin{document}

\newcommand\XXX[1]{{\textcolor{red}{\textbf{x\ #1\ x}}}}

\title{The $\infty$ galaxy: a candidate direct-collapse supermassive black hole between two massive, ringed
nuclei}


\author[0000-0002-8282-9888]{Pieter van Dokkum}
\affiliation{Astronomy Department, Yale University, 219 Prospect St,
New Haven, CT 06511, USA}
\affiliation{Dragonfly Focused Research Organization, 150 Washington Avenue, Suite 201, Santa Fe, NM 87501, USA}
\author[0000-0003-2680-005X]{Gabriel Brammer}
\affiliation{Cosmic Dawn Center (DAWN), Niels Bohr Institute, University of Copenhagen, Jagtvej 128, K\o benhavn
N, DK-2200, Denmark}
\author[0009-0005-2295-7246]{Josephine F.\ W.\ Baggen}
\affiliation{Astronomy Department, Yale University, 219 Prospect St,
New Haven, CT 06511, USA}
\author[0000-0002-7743-2501]{Michael A.\ Keim}
\affiliation{Astronomy Department, Yale University, 219 Prospect St,
New Haven, CT 06511, USA}
\author[0000-0002-5554-8896]{Priyamvada Natarajan}
\affiliation{Astronomy Department, Yale University, 219 Prospect St,
New Haven, CT 06511, USA}
\author[0000-0002-7075-9931]{Imad Pasha}
\affiliation{Astronomy Department, Yale University, 219 Prospect St,
New Haven, CT 06511, USA}
\affiliation{Dragonfly Focused Research Organization, 150 Washington Avenue, Suite 201, Santa Fe, NM 87501, USA}

\begin{abstract}

We report the discovery of an unusual $z=1.14$ object,
dubbed the $\infty$ galaxy, in JWST imaging of the COSMOS field.
Its rest-frame near-IR light is dominated by two compact
nuclei with stellar masses of
$\sim 10^{11}$\,\msun\ and a projected separation of $10$\,kpc. Both nuclei have
a prominent ring or shell around them, giving the galaxy the appearance of a figure eight or
an $\infty$ symbol.  The morphology 
resembles that of the nearby system II\,Hz\,4, 
where the head-on collision of two galaxies with parallel disks
led to the formation of collisional rings around both of their bulges.
Keck spectroscopy,
VLA radio data, and Chandra X-ray data show that the $\infty$ galaxy hosts an 
actively accreting supermassive black hole (SMBH), with quasar-like radio and X-ray luminosity.
Remarkably, the SMBH is not associated with either of the two nuclei, but is in between them in both position
and radial velocity. Furthermore,
from excess emission in the NIRCAM F150W filter we infer that the SMBH
is embedded in an extended distribution of H$\alpha$-emitting gas, with a rest-frame equivalent
width ranging from 400\,\AA\ -- 2000\,\AA.
The gas  spans the entire width
of the system and was likely shocked and compressed at the
collision site, in a galaxy-scale equivalent of what
happened in the bullet cluster.
We suggest that the SMBH formed within this gas in the immediate
aftermath of the collision, when it was dense and highly turbulent.
If corroborated with simulations and follow-up JWST spectroscopy, this would demonstrate 
that `direct' SMBH formation by a runaway gravitational collapse is possible
in extreme conditions.

\end{abstract}


\section{Introduction}

The James Webb Space Telescope (JWST) has enabled us to find and study galaxies that were
beyond the reach of previous observatories. This is most obviously the case 
at the highest redshifts, with JWST having found galaxies out to $z\sim 14$ \citep{curtislake:23,carniani:24} and uncovering
previously unknown populations of red galaxies at $z\sim 6-10$ \citep{labbe:23,matthee:24}.
JWST is also providing qualitatively new information at lower redshifts, even for relatively
bright objects, because of its exquisite
spatial resolution, sensitivity, access to wavelengths beyond $2\,\mu$m,
and spectroscopic capabilities \citep[see, e.g.,][for some examples]{mowla:22,nelson:23,beverage:25,jain:25}.

In this context it is worthwile to search for unusual objects in public wide field
JWST surveys, even in areas of the sky
that have been studied extensively in prior HST programs. One of the first such
surveys was COSMOS-Web \citep{casey:23}, a multi-band NIRCAM program
in the COSMOS field \citep{scoville:07}. Any objects found in this area of the sky have a wealth of
data available at other wavelengths, especially in the part of the field that
coincides with the CANDELS survey \citep{koekemoer:11,momcheva:16}.

Two of us (GB and PvD) conducted a search for interesting objects in the COSMOS-Web NIRCAM
data, by visually inspecting a mosaic of the reduced data.\footnote{https://github.com/gbrammer/grizli-notebooks/tree/main/JWST}
Several objects stood out to us. The first was a complete Einstein ring around a
compact, massive galaxy at $z\approx 2$, presented in
\citet{dokkum:24lens}. The Einstein ring was independently identified by the COSMOS-Web
team \citep{mercier:24}.

\begin{figure*}[ht]
  \begin{center}
  \includegraphics[width=0.98\linewidth]{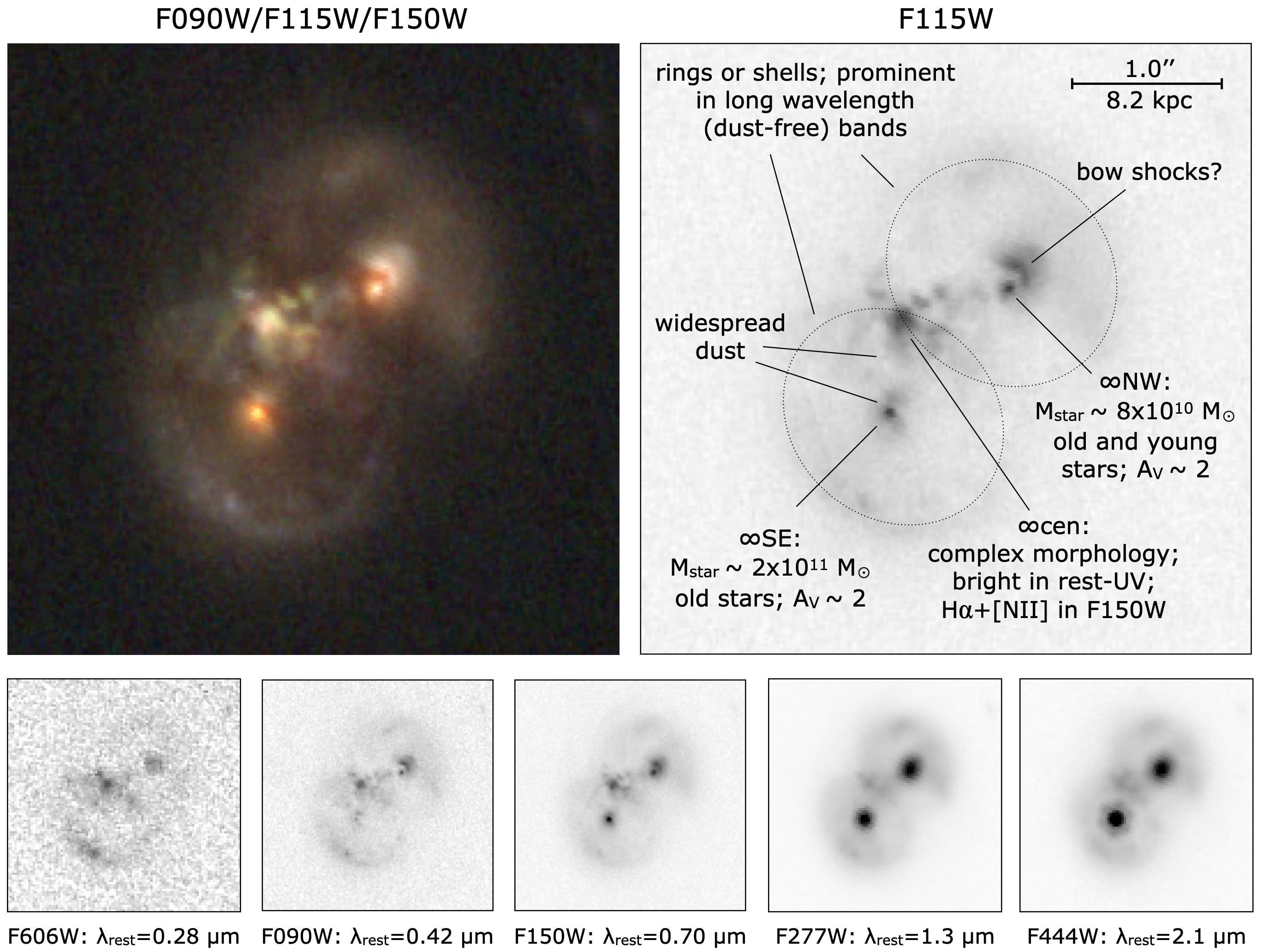}
  \end{center}
\vspace{-0.2cm}
    \caption{
JWST and HST images of the $\infty$ galaxy.
The top left panel shows a color image created from JWST/NIRCAM F090W, F115W,
and F150W data, sampled at $0\farcs 02$\,pix$^{-1}$.
Key morphological elements are indicated in the F115W rendition at top right. 
The row of small panels at the bottom show the appearance of the object in
selected HST/ACS and JWST/NIRCAM bands. In the rest-frame $K$ band
the flux is dominated by two bright and compact nuclei, each with an
apparent ring or shell around it. 
}
\label{jwst_im.fig}
\end{figure*}

The second is shown in Fig.\ \ref{jwst_im.fig}, and the subject of this Letter. It has
a highly unusual
morphology: two very compact, red nuclei, each surrounded by a conspicuous ring or shell.\footnote{We initially thought
that the $\infty$ galaxy was a single galaxy with a ring on the edges of two adjacent exposures, and that an astrometry
problem had rendered it twice in the mosaic.}
We nicknamed the object the $\infty$ galaxy because of its dual-ring morphology
at rest-frame optical wavelengths. 
Follow-up observations, detailed below, showed that the
galaxy is at $z=1.14$, and
that it hosts an active supermassive black hole (SMBH) in between the two nuclei.

\section{Data}

\subsection{Description of the HST and JWST Imaging}

\begin{figure}[htb]
  \begin{center}
  \includegraphics[width=1.0\linewidth]{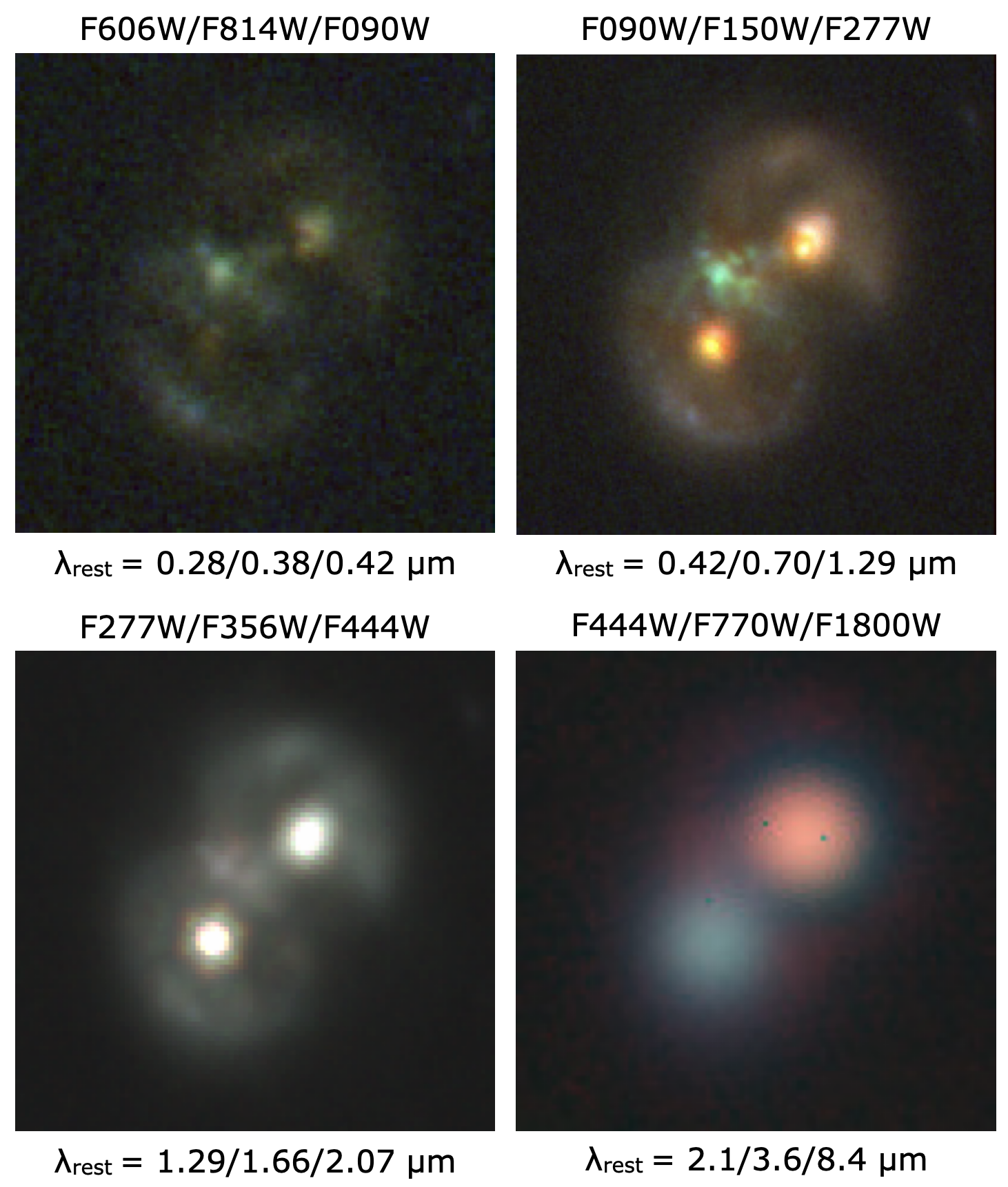}
  \end{center}
    \vspace{-0.3truecm}  
    \caption{\small 
{\em Top left:} HST/ACS F606W, F814W, and JWST/NIRCAM F090W, with the F090W
image smoothed to the HST resolution. 
{\em Top right:}
JWST/NIRCAM F090W, F150W, and F277W. {\em Bottom left:} JWST/NIRCAM
F277W, F356W, and F444W. {\rm Bottom right:} JWST/NIRCAM F444W, JWST/MIRI F770W,
and JWST/MIRI F1800W, with F444W and F770W smoothed to the F1800W resolution.
   }
   \label{morephot.fig}
    \vspace{-1pt}
\end{figure}

The $\infty$ galaxy is at \RA{10}{00}{14.189}, \DEC{+}{2}{13}{11.67} 
(J2000) in the CANDELS/3D-HST area of the COSMOS field \citep{momcheva:16}.
We reduced JWST NIRCAM and MIRI imaging 
from the Cosmos-Web \citep{casey:23} and PRIMER \citep{donnan:24} surveys with
the imaging module of the {\tt grizli} code \citep{grizli}, sampled at
a resolution of $0\farcs 02$\,pix$^{-1}$
(NIRCAM, SW) or $0\farcs 04$\,pix$^{-1}$ (NIRCAM, LW and MIRI).
The small panels of Fig.\ \ref{jwst_im.fig} show single-band images of the galaxy at different
wavelengths. RGB color images ranging from HST/ACS F606W all the way to JWST/MIRI F1800W
are shown in Fig.\ \ref{morephot.fig}.

The F277W, F356W, and F444W filters 
sample the rest-frame near-IR and are not very sensitive to dust, emission
lines, or young stars.  The nuclei and the rings are prominent in these
filters, with little color variation between them. 
This shows that the rings are stellar, and that their
separation from the nuclei at shorter wavelengths
is not caused by radial gradients in dust extinction.
In between the rings is a compact region that is bright in the rest-frame UV and in several JWST bands. It is particularly bright in F150W, which contains the redshifted H$\alpha$, [N\,II], and [S\,II] lines.
This excess F150W emission is conpicuous in the F090W/F150W/F277W RGB image as it shows
up as green. The NW nucleus is relatively bright in the 
MIRI MRS F1800W filter. This filter samples the rest-frame
$8\,\mu$m PAH emission band.

\subsection{Keck Spectroscopy}

\subsubsection{Observations and Data Reduction}

The $\infty$ galaxy was observed with LRIS on the Keck\,I telescope on three occasions,
April 24 2023, November 7 2024, and November 27 2024. For all observations we used
the 300\,lines\,mm$^{-1}$ grism, blazed at 5000\,\AA, in the blue and the 6800\,\AA\
dichroic. Other settings are listed in Table \ref{tab:observations}. 
The 2023 data are relatively shallow but they cover a large wavelength range, and include
the important redshifted H$\beta$ and [O\,III] lines.

\begin{table}
\begin{small}
    \centering
    \caption{LRIS observations}
    \begin{tabular}{cccccc}
        \hline
        \hline
        Date & Grism & Grating & $t_{\rm exp}$ [s] & PA [$^{\circ}$] \\
        \hline
        4/24/2023 & 300/5000 & 400/8500 & 1800 & $-60$ \\
        11/7/2024 & 300/5000 & 600/10000 & 4500 & $-17$ \\
        11/27/2024 & 300/5000 & 600/10000 & 8400 & $-45$ \\
        \hline
    \end{tabular}
    \label{tab:observations}
    \end{small}
\end{table}

The data were reduced using a combination of standard packages and custom techniques.
We used {\tt PypeIt} \citep{pipeit}
to generate a sky model, identify cosmic rays, and fit an approximate wavelength solution for each
frame.
The code was run in the AB subtraction mode for the red side data of 2024 November 28, to optimize sky subtraction. All other data were reduced in the normal reduction mode. This served as the basis for tailored reduction steps.
We used sky emission lines to correct the wavelength solutions in all
individual frames.
This was particularly important for the data beyond $1\,\mu$m in 2023, where
the standard reduction produced significant errors.
The spectra were then placed on a common wavelength grid.
Next, the 2D spectra were corrected for s-distortion by determining the
location of the spectral trace as a function of wavelength, and
shifting the spectra in the spatial direction.
Finally, a master frame was created as a weighted combination of individual frames. We determined weights by summing across the wavelength direction and fitting the trace of the brightest object with a Gaussian, taking the ratio of the integrated flux to the FWHM as that frame's weight. Cosmic ray masks, reduced together with science frames using the same steps as previously described so that all affected pixels were masked, were utilized in the sum so that the master frame was the weighted average of all non-masked pixels.

\begin{figure*}[ht]
  \begin{center}
  \includegraphics[width=0.95\linewidth]{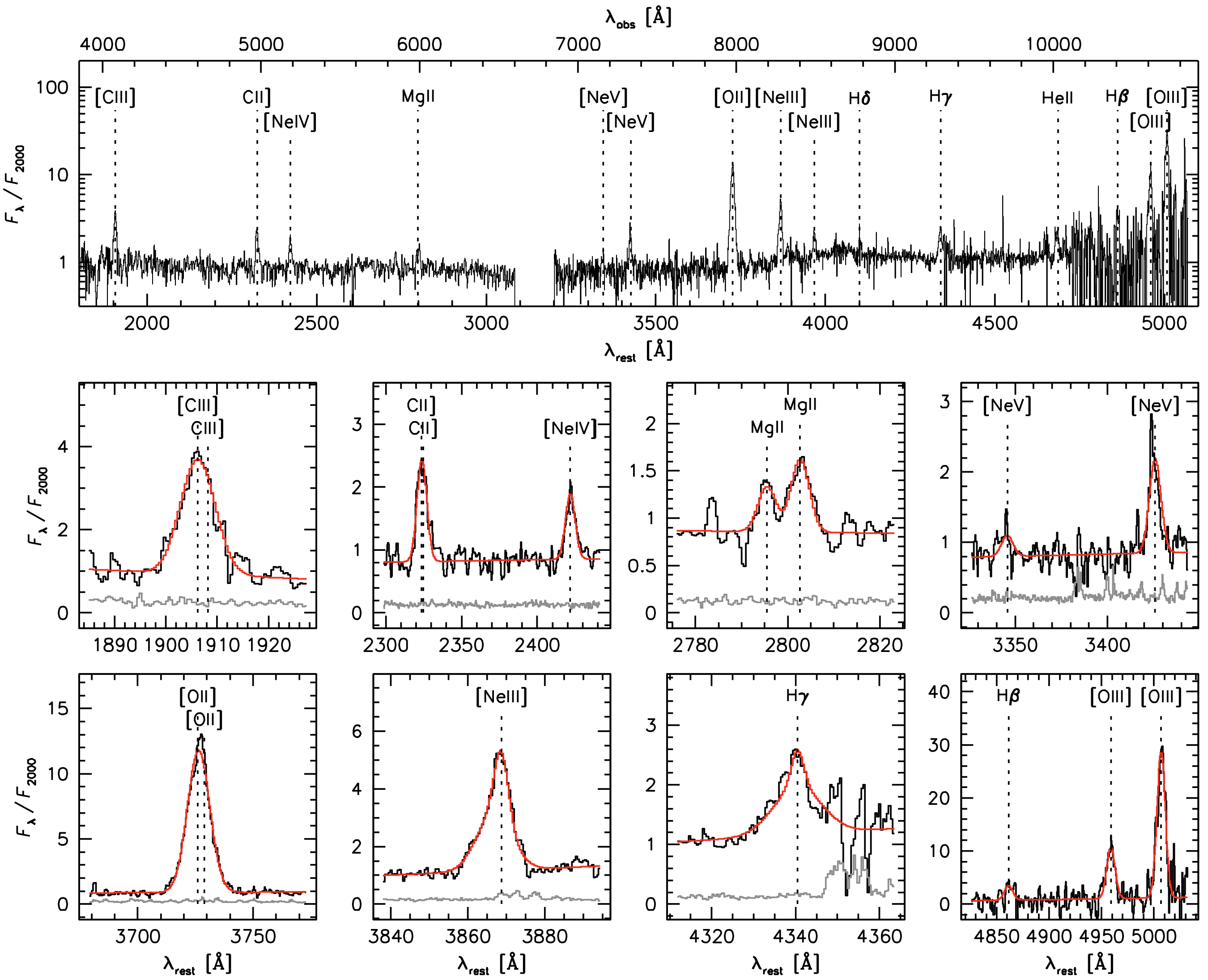}
  \end{center}
    \vspace{-0.3truecm}  
    \caption{\small 
Combined Keck/LRIS spectrum of the $\infty$ galaxy. The top
panel shows the entire wavelength range, with a logarithmic $y$-axis to capture the
large dynamic range of the emission lines. Zoomed views are on
a linear scale.
Black lines are
observations, grey lines indicate the errors, and
red lines are Gaussian emission line fits
redshifted to $z=1.1403$. For [Ne\,III] and H$\gamma$ a
second component was fit to model the outflow and broad component respectively.
The spectrum shows a high degree of ionization and resembles that of
a narrow line AGN.
   }
   \label{spectrum.fig}
\end{figure*}

Prior to combining the 2023 and 2024 data we performed a relative flux
calibration. In 2023 no
standard star was observed.
We obtained a wavelength-dependent flux calibration by
fitting the strength of OH emission lines from the night sky to a theoretical spectrum.
For the 2024 data 
we used observations of the standard
star Feige 34 to derive the response function of the instrument. 
The three datasets were combined by requiring that the average
fluxes at $6000$\,\AA\,$<\lambda<6200$\,\AA\ (blue) and
$7000$\,\AA\,$<\lambda<7200$\,\AA\ (red) are the same for the
three datasets. The shallow
2023 data were only used in the far red, where there is no 2024 coverage.
Absolute calibration was performed in the following way.
We measured the through-the-slit flux in
the HST/ACS F814W image, after convolution with a FWHM\,=\,$1\arcsec$
Gaussian. We then integrated the LRIS spectrum
over the wavelength range of the F814W filter.
The ratio of the F814W flux and the integrated spectral flux
gives the calibration factor to
bring the spectrum to a calibrated flux density scale.

The combined spectrum is shown in Fig.\ \ref{spectrum.fig}. It was extracted from a $1\farcs 1$ region
centered on the peak of the emission line flux in the spatial direction.
The spectrum is characterized by many strong emission lines, including high ionization species
such as [Ne\,V]\,$\lambda\lambda 3346,3426$. Remarkably, the reddest detected line ([O\,III]\,$\lambda 5007$) is at 
an observed wavelength of $10,716$\,\AA, where LRIS has very low sensitivity.

\subsubsection{Emission Line Fits}
\label{linefit.sec}

Most of the emission lines are fit with a combination of a Gaussian and a linear function.
Inputs are the observed spectrum and an error spectrum. The error spectrum is
determined by extracting spectra from empty regions away from the galaxy, and
determining the rms scatter in these spectra as a function of wavelength.
Fits of single lines have four free parameters: the width and normalization of the Gaussian, and
the slope and offset of the linear function. In most cases several
lines are fit simultaneously, as they are close in wavelength; examples
are the [C\,III], C\,III] lines and the Mg\,II doublet. The redshift is held
fixed at $z=1.1403$, as initially determined from a joint fit to all isolated lines.
Most of the lines have widths of $\sigma \approx 250$\,\kms. 

For two lines we fit more complex functions. The [Ne\,III]\,$\lambda3869$ line has a clear
blue wing that we approximate by fitting a second Gaussian. This second Gaussian has
a central wavelength that is blueshifted by $\approx 150$\,\kms\ and a width
$\sigma\approx 350$\,\kms. The other line is H$\gamma$, the only Balmer line that has
a reasonably high S/N ratio. The line appears to have a broad
component, although the interpretation is hampered by the fact that the
red side of the line suffers from contamination by sky lines. Fitting two Gaussians with
fixed centers gives widths of $\sigma = 104 \pm 26$\,\kms\ for the narrow component
and $\sigma = 397 \pm 45$\,\kms\ for the broad component. The broad component's
flux is $\sim 5\times$ higher than that of the narrow component. We note here
that H$\alpha$ is likely extremely bright (see Fig.\ \ref{morephot.fig}),
but is not accessible to ground-based spectroscopy as it falls in the H$_2$O absorption
gap between the $J$ and $H$ bands.

\subsection{Additional Data}

\subsubsection{X-ray Luminosity and Hardness Ratio}

The X-ray luminosity is determined from Chandra observations, obtained from the
IPAC COSMOS archive \citep{civano:16}. The galaxy is a strong detection with $74\pm 10$ counts in the
$0.5-7$\,keV band for an exposure time of 16.2\,ks. The corresponding
flux is $F_{0.5-7\,{\rm keV}} =(4.1\pm 0.6)
\times 10^{-14}$\,erg\,s$^{-1}$\,cm$^{-2}$, where we used the Cycle 8 sensitivity of the ACIS
detector. Assuming a powerlaw spectrum with $\Gamma=1.8$ (see below) the
luminosity
in the rest-frame $2-10$\,keV band is 
$L_{\rm X} \approx 1.5 \times 10^{44}$\,erg\,s$^{-1}$.

With $54\pm 8$ counts in the $0.5-2$\,keV band and $19\pm 5$ counts in the $2-7$\,keV band
the hardness ratio, ${\rm HR} = ({\rm H}-{\rm S})/({\rm H} + {\rm S}) = -0.48 \pm 0.12$,
indicating a relatively soft spectrum.
The spectral slope, calculated from the fluxes in the soft and hard band,
is $\Gamma \approx 1.8$.

\subsubsection{Radio Observations}

The $\infty$ galaxy is a very strong radio source. VLA observations were
obtained from the IPAC COSMOS archive.
The galaxy has flux densities of $S_{\nu}=1.5$\,mJy at 1.4\,GHz and $S_{\nu}=0.43$\,mJy at 3\,GHz.
We also obtained data from LOFAR \citep[the Low-Frequency Array;][]{lofar} at 144\,MHz, and measure a
flux density of $S_{\nu} = 23.4$\,mJy. We obtain a rest-frame radio luminosity of $L_{\rm 144\,MHz} =
2\times 10^{26}$\,W\,Hz$^{-1}$, where we used a spectral index of $\alpha=1.3$ ($S_{\nu}\propto \nu^{-\alpha}$).
The spectral index is determined from
a fit to the $144$\,MHz, 1.4\,GHz, and 3\,GHz flux densities. 

Below we use the centroid of the VLA 3\,GHz observation to pinpoint the location
of the SMBH within the JWST NIRCAM images. We verify the relative astrometry
between the JWST and VLA data in the following way. First, we obtained VLA 3\,GHz and
HST/ACS F814W data from the IPAC archive in a $4' \times 4'$ field centered
on the $\infty$ galaxy. Sources are detected in both images using
SExtractor \citep{bertin:96}, and the centroids of these sources are
compared after cross-matching the catalogs. 
We find nine sources that are bright and compact in the F814W image and in the VLA data
(besides the $\infty$ galaxy itself). There is a small systematic
offset in position, of $-0\farcs 11\pm 0\farcs 02$ in RA and $-0\farcs 01 \pm 0.02$ in DEC.
Next, we compare the position of the $\infty$ galaxy in the IPAC F814W image to that in
our reductions. Our reductions show the same offset with respect to the
IPAC F814W data as the VLA data do, 
and we conclude that the VLA 3\,GHz map is on the same astrometric system as our
HST and JWST reductions, with an accuracy of $\lesssim 0\farcs04$.

\section{Analysis}

\subsection{Stellar Populations and Sizes of the Two Nuclei}

\begin{figure}[htb]
  \begin{center}
  \includegraphics[width=1.0\linewidth]{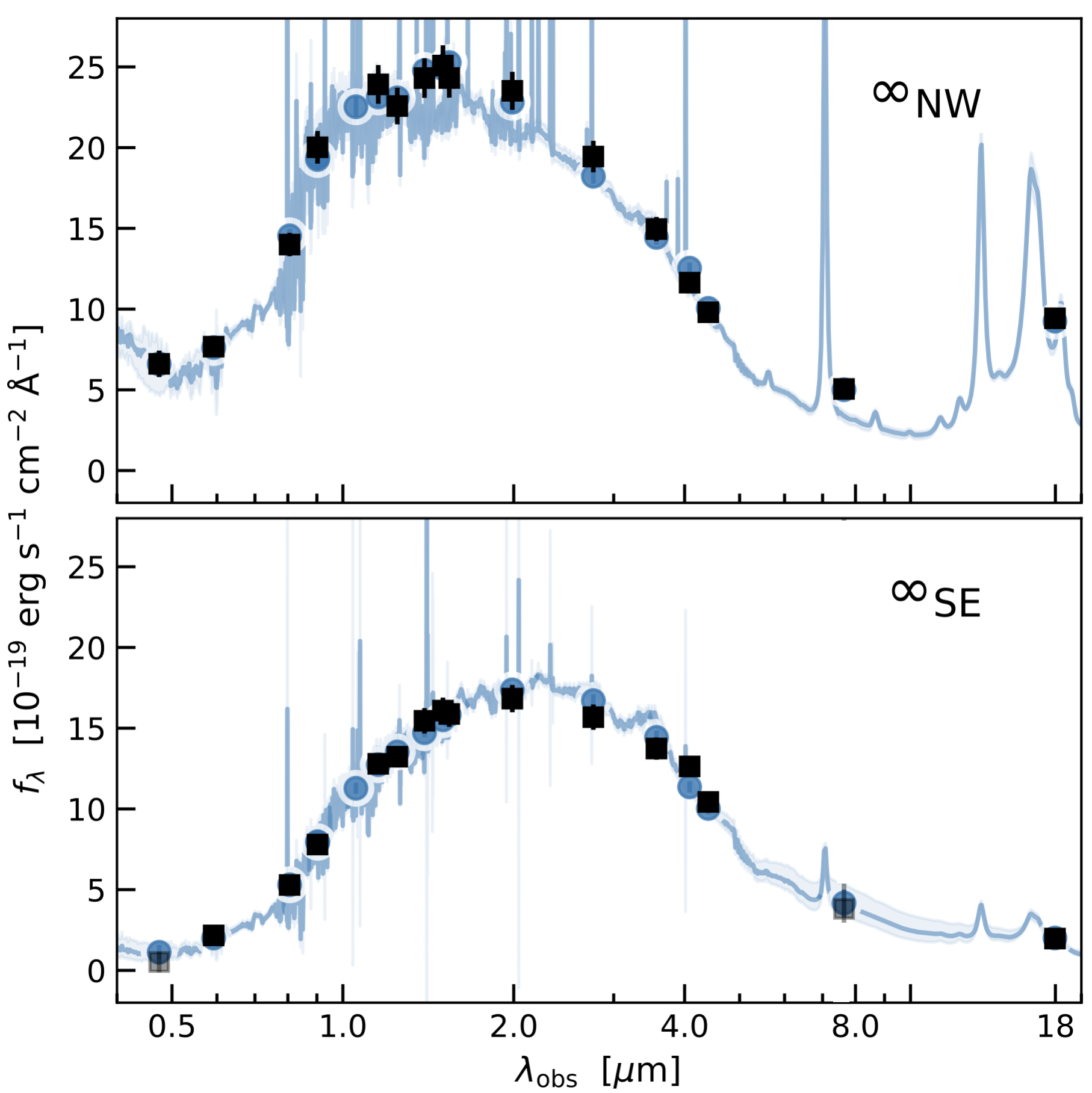}
  \end{center}
    \vspace{-0.3truecm}  
    \caption{\small  
{\em Top:} Stellar population fit, using the {\tt eazy} code, for $\infty$\,NW.
The implied stellar mass is $8\times 10^{10}$\,\msun. Note that the
galaxy has strong $8\,\mu$m PAH emission, sampled in the F1800W MIRI band.
{\em Bottom:} Fit for $\infty$SE. This galaxy has a mass of $1.8\times 10^{11}$\,\msun\ and mostly composed of evolved stellar
populations. Both galaxies are dusty, with $A_V \sim 2$.
   }
   \label{ssp.fig}
    \vspace{-1pt}
\end{figure}

The two red nuclei (marked $\infty$\,SE and $\infty$\,NW) dominate the rest-frame near-infrared light.
They are remarkable in their own right, as they are very bright
and also very compact. We performed stellar population fits to determine approximate stellar
masses. Photometry in all available HST and JWST bands was obtained with
an automated pipeline following procedures that have been used for other public JWST imaging surveys
\citep{valentino:23}.  The two nuclei of the $\infty$ system were identified as unique sources in the automatic catalog procedure without modification, and their multiwavelength photometry extracted within
$D=0\farcs5$ circular apertures were fit with the {\tt eazy} code \citep{eazy}, with
the redshift held fixed to $z=1.14$.
The code fits linear combinations of template SEDs to the observed HST and JWST photometry.  The individual templates have a range of star formation histories (with nebular emission from the youngest stellar populations) and dust attenuation;
each template has an associated $M/L$ ratio and the total stellar mass
can be estimated from their weights in the fit \citep{labbe:23}.
The best fitting SEDs for the two nuclei are shown in Fig.\ \ref{ssp.fig}.

\begin{figure*}[t]
  \begin{center}
  \includegraphics[width=1.00\linewidth]{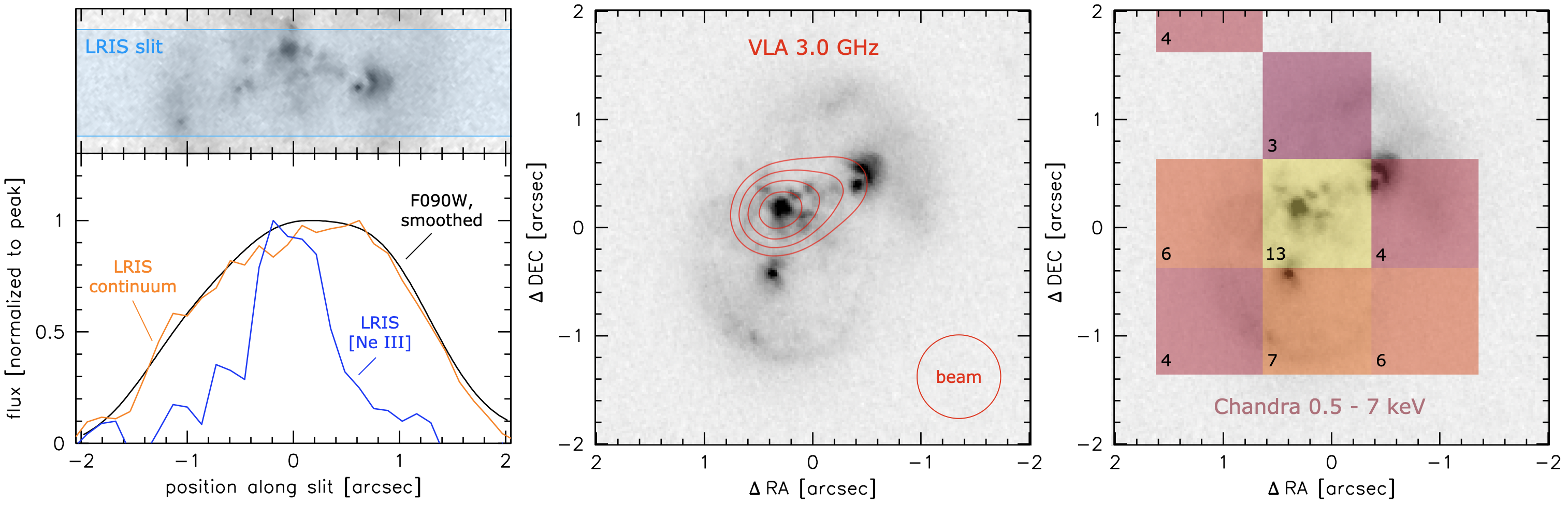}
  \end{center}
    \vspace{-0.3truecm}  
    \caption{\small Localization of the active black hole.
{\em Left panel:} Comparison of the spatial profile of the [Ne\,III] line and the
continuum emission along the LRIS slit. The [Ne\,III] line is unresolved
at the LRIS resolution, and coincides with the compact $\infty$\,cen region
in Fig.\ 1. {\em Middle panel:} VLA 3\,GHz map, with a FWHM beam of $0\farcs 77$. The
strong radio source coincides with $\infty$\,cen to $\lesssim 0\farcs 1$.
{\em Right panel:} The object is also a strong X-ray source, with $L_{\rm X}
\approx 1.5\times 10^{44}$\,erg\,s$^{-1}$. The number of detected photons
is shown in the lower left corner of each
Chandra pixel. The pixel containing the $\infty$\,cen region has the
highest number of photons. For clarity pixels with $\leq 2$ counts are not shown.
   }
   \label{agn.fig}
\end{figure*}

The NW nucleus is best fit by a mixture of evolved stellar populations, young populations,
and ongoing star forming. The F1800W data point requires templates with
strong $8\,\mu$m PAH emission. The mass is $M_{\rm stars} \sim 8\times 10^{10}$\,\msun,
and there is significant dust attenuation with $A_V \sim 2.3$\,mag.
The SE nucleus is even more massive and older, with low levels of ongoing star formation.
We find $M_{\rm stars} \sim 1.8\times 10^{11}$\,\msun. Its dust attenuation is high
for such an old, evolved galaxy, with $A_V \sim 2.1$\,mag. There is
an obvious dust lane across $\infty$SE in the JWST images, and it may be that
some or most of the dust is in its local foreground rather
than within the nucleus itself.

We fit S\'ersic profiles to the F444W images to measure the sizes of the two nuclei, using the GALFIT code \citep{galfit}. Free parameters are the position in the image, the position angle, the effective radius along the major axis ($r_{\rm e}$), the S\'ersic index ($n$), and the projected minor-to-major axis ratio ($b/a$). Circularized effective radii are calculated as $r_{\rm e,c} = r_{\rm e} \sqrt{b/a}$. 
To fit the two nuclei, we mask the surrounding complex emission using the following approach. First, the background is estimated using $\sigma$-clipped statistics with a filter size of 5 pixels and the background RMS. This background is subtracted from the data and then convolved with a 2D Gaussian kernel with a FWHM of 3 pixels.
Next, sources are detected in the convolved, background-subtracted image using a 1$\sigma$ detection threshold, where $\sigma$ corresponds to the background RMS prior to convolution. The resulting detection is used to create a mask map, where non-zero pixels are ignored by GALFIT during the fitting process.

The NW nucleus is well-resolved, with $r_{\rm e, circ}= 0\farcs11$, 
$n=1.2$, and $b/a= 0.8$. With $z=1.1403$ the physical half-light radius
$r_{\rm e, circ}=0.9$\,kpc. The best-fit integrated F444W magnitude is $m_{444}=20.7$. 
We find that the  $\infty$SE nucleus is only just resolved, with $r_{\rm e} = 0\farcs052$,
$b/a=0.9$, and $m_{444}=20.5$. The circularized effective
radius $r_{\rm e, circ}=0.4$\,kpc. 
The effective radius
has a systematic uncertainty of $0.1-0.2$\,kpc, based on a comparison
of results using
WebbPSF and two empirical PSFs \citep[following][]{baggen:24}.

The high stellar masses and small sizes imply that the nuclei are  extraordinarily
compact. The $\infty$SE nucleus, in particular,
is a factor of $\sim 6$ smaller than expected from the $z=1$ size-mass relation of quiescent galaxies \citep{wel:14}.

\subsection{An Active Supermassive Black Hole}
\label{agn.sec}

The optical spectrum is
characterized by strong emission lines that imply a high degree of ionization.
With ${\rm [O\,III]/H}\beta = 10.3 \pm 1.3$ and ${\rm [Ne\,V]/[Ne\,III]} =0.32\pm 0.04$ the object firmly falls in the
AGN region of diagnostic diagrams that separate star formation from nuclear activity \citep[see][]{negus:23,cleri:23}.
Furthermore, the pronounced blue wing of the [Ne\,III] line (see \S\,\ref{linefit.sec})
is commonly interpreted as an AGN-driven outflow in the narrow line region \citep{mullaney:13}.

The interpretation of the line emission as photo-ionization by an actrive SMBH
is strongly supported by the radio and X-ray luminosity of the $\infty$ galaxy.
The rest-frame
radio luminosity of $L_{144\,{\rm MHz}} \sim 2 \times 10^{26}$\,W\,Hz$^{-1}$ is comparable to the most powerful
radio-loud AGNs in the local Universe \citep{morabito:25}.
The spectral index is steep, 
$S_{\nu} \propto \nu^{-1.3}$, an indicator of the early stages of jet evolution in a 
young ($10^{4-5}$\,yr) AGN \citep{odea:98}. 
The X-ray luminosity of
$L_{\rm X}\sim 1.5\times 10^{44}$\,erg\,s$^{-1}$ in the rest-frame $2-10$\,keV band puts it in
the border region between luminous Seyfert galaxies and quasars, and
the spectral slope of $\Gamma \approx 1.8$ is typical for
unobscured or moderately obscured AGNs \citep{corral:11}.

We can obtain a rough estimate of the black hole mass from the width of the H$\gamma$ line.
As discussed in \S\,\ref{linefit.sec}, the line can be fitted with a combination of a narrow component and
a dominant broad component, with
FWHM\,$=940 \pm 110$\,\kms. Combined with the estimated 
H$\alpha$ luminosity of $L_{{\rm H}\alpha} \sim 3~L_{{\rm H}\beta} \sim 3\times 10^{41}$\,erg\,s$^{-1}$,
this gives an estimated black hole mass of $M_{\rm BH} \sim 10^6$\,\msun\
\citep[Eq.\ 6 in][]{greene:05}.
We note that we cannot constrain the red wing of the H$\gamma$ line as it falls on a sky line, and it could be that
the blue wing should be modeled as an outflow rather than the short wavelength half of a symmetric broad component. Furthermore, 
the black hole mass may be significantly lower
if chaotic motions in the surrounding gas contribute to the H$\alpha$ broadening.

Interestingly, the X-ray luminosity is similar to the Eddington luminosity for a $10^6$\,\msun\
black hole ($L_{\rm Edd} \approx 1.3 \times 10^{44}$\,erg\,s$^{-1}$).
However, this is indicative only, given the large uncertainty in the black hole mass.
Furthermore, given the steep spectral index,
the X-ray and radio luminosities could be significantly boosted by non-thermal
emission from a young jet \citep[see][]{odea:98}.

\subsection{Location of the Black Hole}

We now arrive at the most striking and unexpected aspect of the $\infty$ galaxy: the active SMBH
does not appear to be associated with either of the two massive stellar nuclei, as might have been expected, but is
located in the complex region in between them (labeled $\infty$\,cen in
Fig.\ \ref{jwst_im.fig}). This is demonstrated in Fig.\ \ref{agn.fig}, which shows that the high
ionization UV/optical emission lines, the radio emission, and the X-rays all come from this region. The strongest constraint comes from the high spatial resolution VLA 3\,GHz map, whose centroid coincides with the brightest pixels of the $\infty$\,cen region to within $\lesssim 0\farcs1$. We note that there is an extension
in the 3\,GHz map towards the $\infty$NW nucleus; this could be emission from a weak obscured AGN or
obscured star formation in that nucleus \citep{smirnov:24}, or emission from a jet.

We also identify the likely ionizing source within
the $\infty$\,cen region, in the form
of a compact blue object that is most clearly seen in the HST F606W band (see Fig.\ \ref{jwst_im.fig}) at the precise location of the centroid of the 3\,GHz map.
It may reflect emission from the immediate vicinity of the black hole, such as thermal radiation from the accretion
disk or light reprocessed in the broad line region \citep{krolik:99}.

As detailed in Appendix A, we use GALEX imaging to estimate the flux in the rest-frame far-UV and the ionizing photon budget. The $\infty$ galaxy is clearly detected in the NUV channel at $\lambda_{\rm rest} \sim 1080$\,\AA\ with a total AB magnitude of 24.7. The galaxy is not resolved at the GALEX resolution of $\approx 5\arcsec$, but based
on the F606W morphology, as well as ground-based $B$ and $u$ band imaging,
it is likely that a considerable fraction of the rest-frame far-UV flux comes from the compact object. We find that the
ionizing photon rate is sufficient to produce the observed [O\,III] luminosity
of $L_{\rm [OIII]} \approx 1.2 \times 10^{42}$\,erg\,s$^{-1}$ for reasonable photon conversion efficiencies of $0.01-0.1$.
These values are consistent
with typical ranges seen in AGN photoionization models
and observations \citep{netzer:90,baskin:05}.

\begin{figure}[ht]
  \begin{center}
  \includegraphics[width=1.00\linewidth]{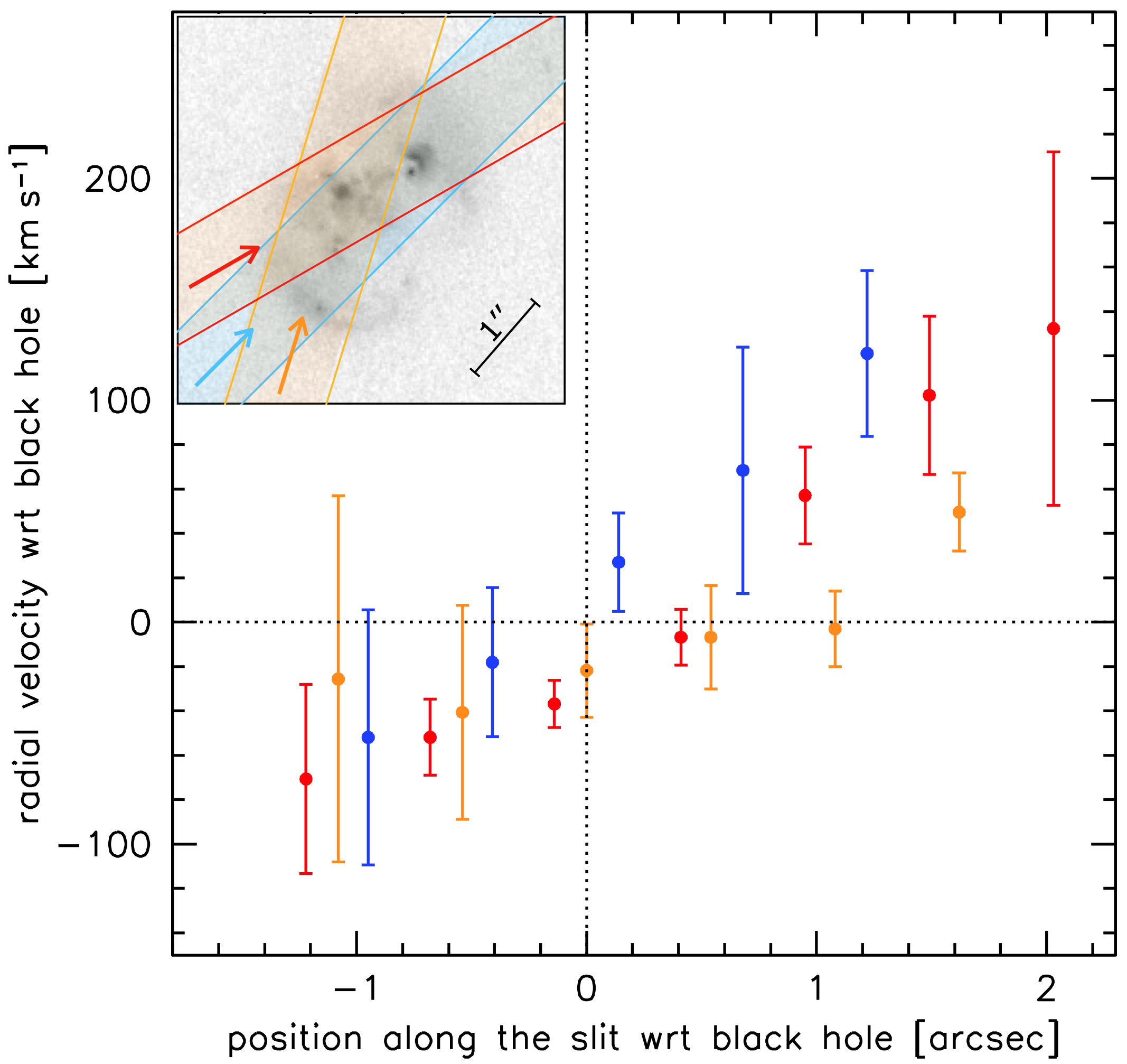}
  \end{center}
    \vspace{-0.3truecm}  
    \caption{\small  
The $\infty$ galaxy was observed three times with Keck/LRIS, with three different slit angles 
(see inset). The radial velocity of the low-ionization [O\,II] doublet is shown as a function
of position along the slit, with the positive $x$-direction indicated by the arrows in the inset. 
There is a gradient of $\pm 100$\,\kms\ that is symmetric with respect to the velocity of the 
highly ionized lines from the SMBH's narrow line region. Note that the velocity profile 
is heavily smoothed by the $1\arcsec$ seeing of the observations.
The intrinsic profile could be a step function, with each side moving at a constant velocity.}
   \label{kinematics.fig}
\end{figure}

\subsection{Radial Velocity of the Black Hole}

The SMBH is not only spatially near the center of the $\infty$ galaxy, but also kinematically. It is difficult to measure radial velocities of the individual components of the galaxy from our ground-based slit spectra given their spatial
resolution of ${\rm FWHM}\approx1\farcs 0$. Nevertheless, while the spatial profiles of
high-ionization lines such as [Ne\,III] are consistent with a point source, the [O\,II] doublet is just-resolved in our Keck spectra. The velocity profile along the slit is shown in Fig.\ \ref{kinematics.fig}. As seen in the inset, the [O\,II] emission that is $1\arcsec -2\arcsec$ away from the black hole coincides with the ring structures, and probably reflects on-going star formation in those regions. There is a gradient, from
approximately $-100$\,\kms\  in the SE
to $+100$\,\kms\ in the NW. The zero-point of the velocity scale is the redshift of the high ionization lines from the
SMBH, and we infer that its radial velocity lies in between that of the two rings.

\begin{figure*}[htb]
  \begin{center}
  \includegraphics[width=1.0\linewidth]{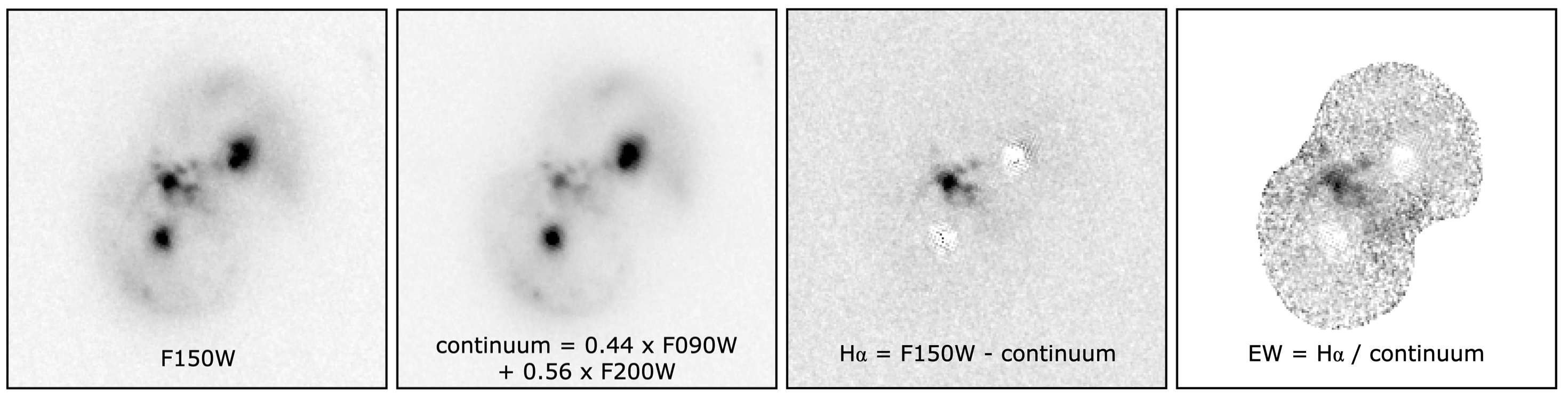}
  \end{center}
    \vspace{-0.3truecm}  
    \caption{\small 
    {\em Left:} JWST F150W image, containing both continuum and the H$\alpha$,
    N\,II, and S\,II emission lines. {\em Second from left:} Continuum image, created
    by interpolating the F090W and F200W images. {\em Second from right:} Emission line
    map, created by subtracting the continuum image from the F150W image.
    {\em Right:} Equivalent width map: division of the emission line
    map by the continuum image.
   }
   \label{halpha.fig}
    \vspace{10pt}
\end{figure*}

\begin{figure}[ht]
  \begin{center}
  \includegraphics[width=1.00\linewidth]{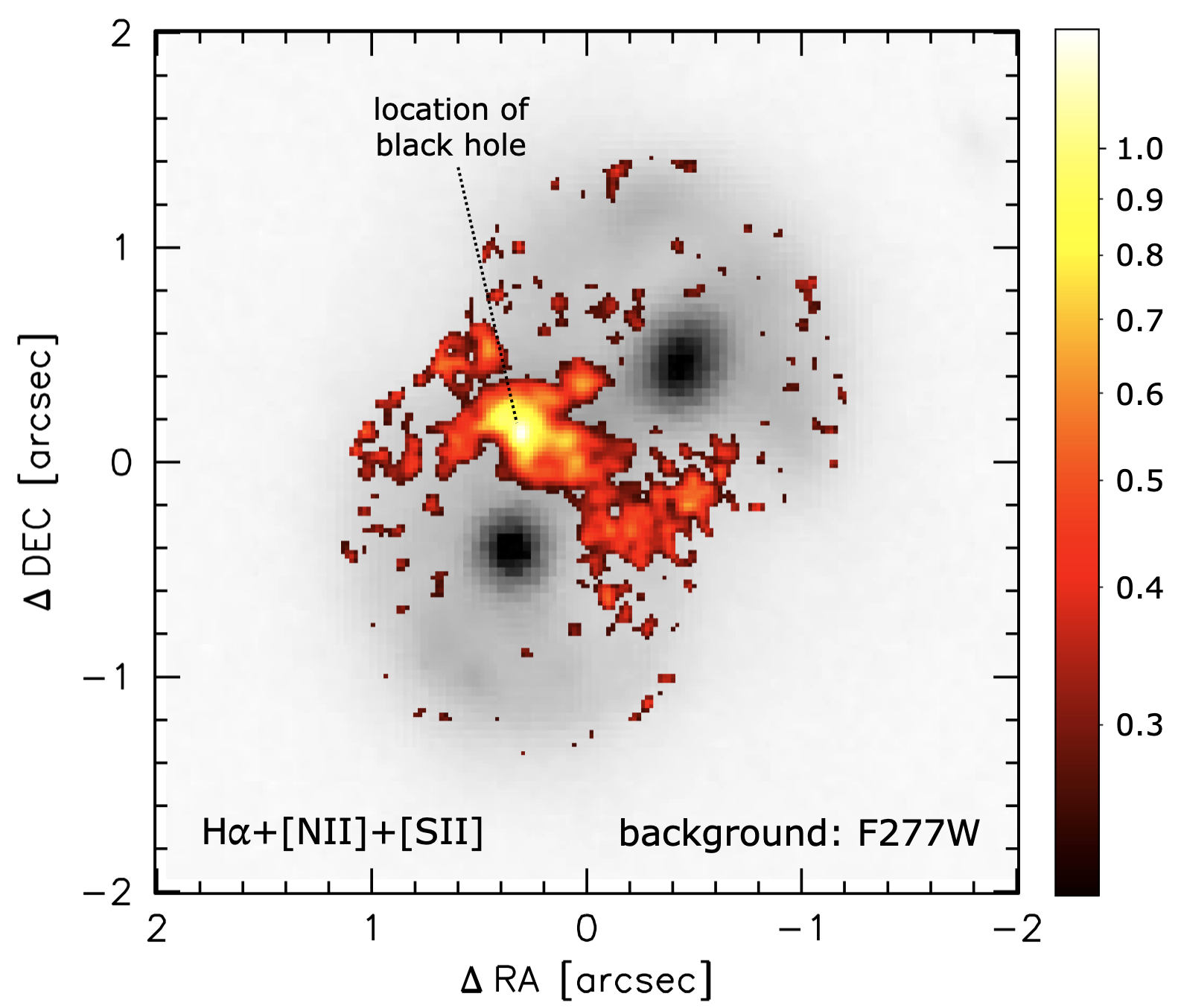}
  \end{center}
    \vspace{-0.3truecm}  
    \caption{\small 
Equivalent width map of the H$\alpha$\,+\,[N\,II]\,+\,[S\,II] line emission in the NIRCAM F150W filter, overplotted on
the F277W (rest-frame $J$) image. The scale bar indicates the strength
of the line emission, expressed as a fraction
of the F150W continuum. The rest-frame equivalent width is
approximately ${\rm EW}_{\rm rest} \approx f \times 1350$\,\AA,
with $f$ the continuum
fraction. The strong emission line source associated with
the black hole is embedded in an elongated ionized
gas distribution with roughly constant EW that is
in between the two nuclei.
   }
   \label{halpha_overlay.fig}
\end{figure}

\subsection{Evidence for an Extended Ionized Gas Distribution Between the Nuclei}

The $\infty$\,cen region has a conspicuous green color in the F090W/F150W/F277W
rendition of Fig.\ \ref{morephot.fig}. This can be attributed to the fact that
the rest-frame H$\alpha$, [N\,II], and [S\,II] emission lines fall in the F150W
filter. We created a continuum image by interpolating between the F090W and F200W
filters, ${\rm F150W}_{\rm cont} = 0.44\times {\rm F090W} + 0.56\times {\rm F200W}$.
Subtracting the continuum from the F150W image shows the complex
line emission in the $\infty$\,cen region (Fig.\ \ref{halpha.fig}).
There is a clear peak with a radial fall-off, as expected for photo-ionization
by a point source.

The right-most panel of
Fig.\ \ref{halpha.fig} shows the ratio
between the line emission map and the continuum, that is,
$f \equiv ({\rm F150W} - {\rm F150W}_{\rm cont})/{\rm F150W}_{\rm cont}$. 
This ratio corresponds to an equivalent width, such that
${\rm EW}_{\rm rest} \approx f \times 1350$\,\AA. Here ${\rm EW}_{\rm rest}$
is the rest-frame equivalent width of the sum of H$\alpha$, [N\,II], and
[S\,II].
In Fig.\ \ref{halpha_overlay.fig} we show the same map, smoothed
by a Gaussian with $\sigma=0\farcs 02$ and overplotted
on the F277W image of the galaxy.

The map shows the luminous and complex $\sim 2$\,kpc-sized region around
the black hole, with ${\rm EW}_{\rm rest} \sim 1500$\,\AA. Furthermore, there is
line emission with ${\rm EW}_{\rm rest} \sim 400$\,\AA\ in an elongated, $\sim 10$\,kpc-sized
structure between the two nuclei.
The ionized gas distribution is perpendicular to the line between the two nuclei, and
the SMBH is embedded within it.
We note that, at present, we do not have information on the dynamics or line ratios within the
extended ionized gas.

\section{Discussion}

The $\infty$ galaxy presents us with not one but two highly unusual
observational characteristics. First, we have two massive, extremely compact, red nuclei surrounded by rings or shells; and second, we have a supermassive black hole with quasar-like radio and X-ray luminosity sitting in between them.

\subsection{What Shaped the $\infty$ Galaxy?}
\label{bullet.sec}

Turning first to the overall morphology, the
$\infty$ system resembles a pair of collisional ring galaxies, 
where the head-on impact of a compact galaxy with a
disk leads to the sweeping up of disk stars
into an outwardly expanding ring.
These rings are mostly composed of pre-existing disk stars
that are herded into similar orbits, combined with young stars that
are formed from compressed gas \citep[see, e.g.,]{lynds:76,appleton:87,hernquist:93,smith:12}.

We suggest that $\infty$NW
and $\infty$SE were originally compact, massive bulges
that served as mutual impactors in
a symmetric collision of two galaxies with disks. The
disks would have to have had a nearly face-on orientation with respect to each other.
Interestingly, the first collisional ring galaxy that was modeled with
N-body techniques,
II\,Hz\,4, has a similar binary morphology \citep{lynds:76}.
It was successfully modeled as a mutual interaction of two disk galaxies,
lending support to our interpretation (see \S\,\ref{binary.sec}).
Arp 147 is another local example
of collisional
ring formation around both the impactor and the impacted galaxy, although
it does not show the striking symmetry of II\,Hz\,4 or the $\infty$ galaxy \citep{gerber:92}.

\begin{figure*}[t]
  \begin{center}
  \includegraphics[width=1.00\linewidth]{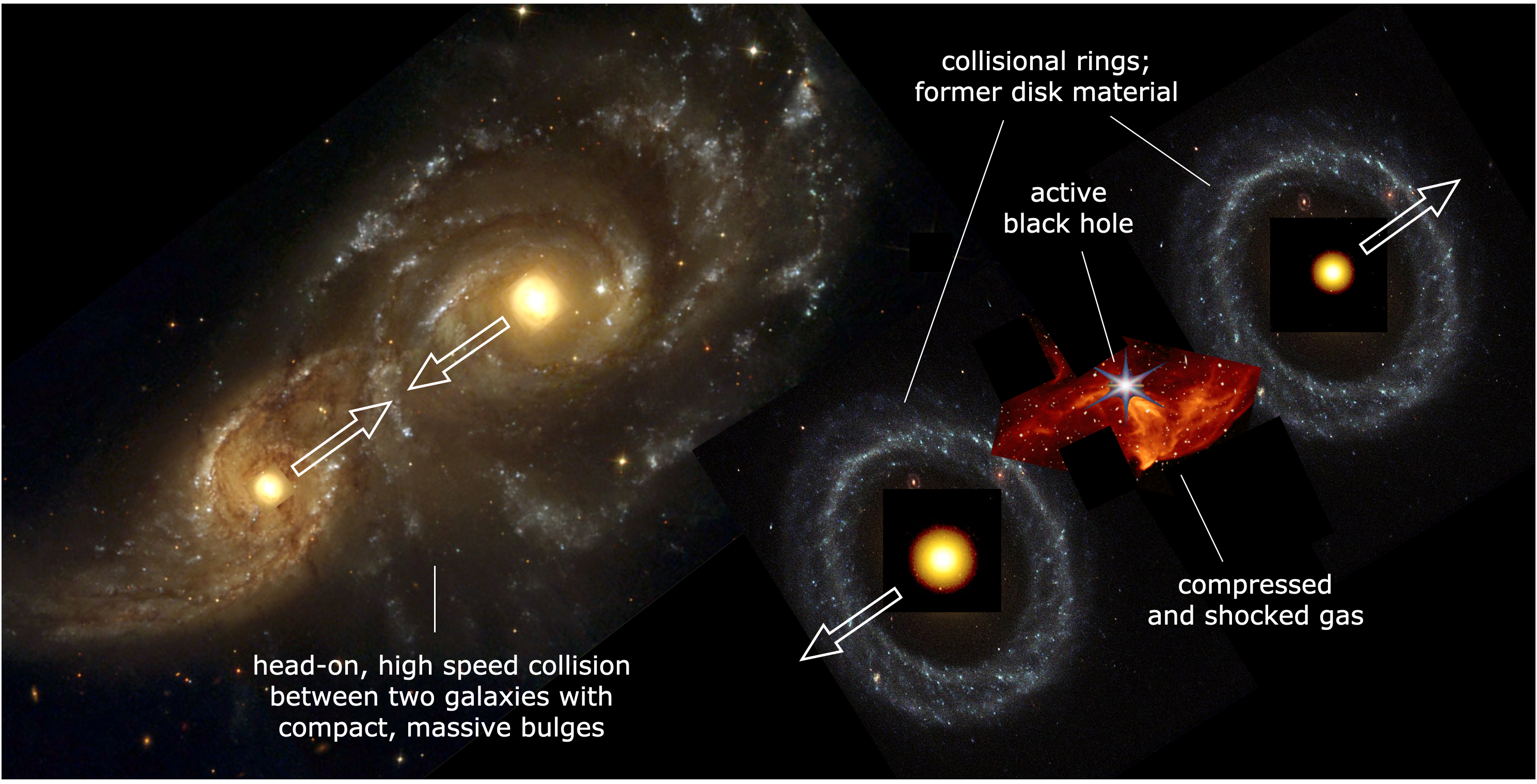}
  \end{center}
    \vspace{-0.3truecm}  
    \caption{\small 
The $\infty$ galaxy is interpreted as the aftermath
of a nearly head-on collision between two face-on disk galaxies with massive, compact
bulges. The bulges survived the collision, and the inner
disk stars were swept up in outwardly expanding 
collisional rings around the bulges. The nearby galaxy II\,Hz\,4 is the prototype for this
kind of binary ring formation \citep{lynds:76}. Compression and shocks in the colliding gas
likely produced a dense gaseous remnant in between the nuclei,
as has been observed in the bullet cluster on much larger scales. It is
proposed that the black hole formed within this gas.
   }
   \label{formation.fig}
\end{figure*}

The presence of the filamentary gas between the nuclei finds a natural
explanation in collision scenarios.
The gas in the pre-existing galaxies  experienced intense shocks and
compression at the time of impact, and could have partially separated
from the stars and dark matter. This process is famously observed in the bullet 
cluster \citep{clowe:06}, and `mini-bullet' scenarios have been proposed to explain 
the formation of dark matter-free dwarf galaxies \citep{silk:19,lee:21,dokkum:22}.
The SMBH is almost certainly responsible for ionizing the gas; the main source
is likely photo-ionization from its UV emission, perhaps augmented by shocks from a jet or a wind
\citep[e.g.,][]{harrison:15}. In this scenario there
are not many stars associated with the gas, which explains the 
extreme equivalent width that is seen in Fig.\ \ref{halpha_overlay.fig}.

\subsection{Origin of the Supermassive Black Hole}

We now focus on the SMBH that is embedded within the ionized gas in between the
two nuclei.

\subsubsection{A Third Galaxy?}

We first consider the possibility that the black hole is in a third galaxy that is unrelated to the two nuclei. 
Inspection of the JWST images shows no obvious candidate at the location of the SMBH: the morphology of the $\infty$\,cen
region is highly complex, with tendrils of emission extending several kpc outwards from the black hole, and there
is no evidence for a bulge, a disk, or other galaxy-like features.
Furthermore, the extreme equivalent width of the H$\alpha$+[N\,II]+[S\,II] complex
implies that any galaxy light must be unusually faint. The rest-frame EW is $\sim 2000$\,\AA\ at the location of the black hole and remains $\gtrsim 1000$\,\AA\ out to a radius of $2-3$\,kpc.  In contrast,
typical AGNs \citep[such as those in the MANGA survey;][]{decontomachado:22} have ${\rm EW}\lesssim 100$\,\AA, due to the significant contribution of the host galaxy's light to the continuum. 

The high radio and X-ray luminosities are difficult to reconcile with a faint dwarf galaxy, 
as it is highly unusual to see quasar-like activity in
a low mass galaxy \citep[see][]{kauffmann:03agn,delvecchio:22}.
Nevertheless,  the 
relations between AGN luminosity and galaxy mass, and
between inferred black hole mass and galaxy mass, have
considerable scatter \citep{harikane:23,mezcua:24}, and
it is of course difficult to exclude the presence of a low mass host in the
$\infty$\,cen region that is outshone by the SMBH.

\subsubsection{A SMBH that Separated from its Host Galaxy?}

Another possible
origin for the SMBH is that it was in the center of a galaxy but separated from it,
either through ejection from one of the two nuclei
or the stripping of an infalling galaxy. Ejections
can occur in a variety of dynamical situations \citep[e.g.,][]{campanelli:07}. In binary black hole mergers with unequal masses and misaligned spins large recoil velocities are typically produced due to gravitational radiation, leading to the possible ejection of the final merged black hole \citep{lousto:11}. Another mechanism for escape is a three-body interaction, when a newly acquired
black hole interacts with a pre-existing binary black hole. Since galaxies
merge frequently and binary black holes are
thought to be long-lived, such escapes may be fairly common \citep{bekenstein:73,saslaw:74,hoffman:07}. Candidates for escaped black holes include the CID-42 double X-ray source in the COSMOS field \citep{civano:10}; a candidate wake of a runaway black hole in the circumgalactic medium of a $z=0.97$ galaxy \citep{dokkumbh:23}; and candidates from the Pan-STARRS1 3$\pi$ Survey \citep{uppal:24}.

Another channel for producing isolated SMBHs is delayed merging.
In some cosmological simulations there is a delay between the infall and final disruption of galaxies and
the infall of their SMBHs, producing isolated SMBHs that are typically close to the center of their
future host galaxies \citep{tremmel:18}. It
has been suggested that these `wandering' black holes can be luminous X-ray sources \citep{ricarte:21},
and the SMBH may be the remnant of on its way to a merger with one of the nuclei.


\subsubsection{In-situ Formation?}
\label{insitu.sec}

The scenarios that we discussed so far can, in principle, occur in any galaxy, and do not predict a close association
between the properties of the SMBH and the (former or future) host galaxy. In both scenarios, the unusual morphology
of the $\infty$ galaxy, the fact that the radial velocity of the SMBH is in between those of the two rings, and
its position halfway between the nuclei in an extended ionized gas distribution are all coincidental. 
Here we discuss a possible origin for the SMBH that is, instead, a {\em consequence} of the unusual
morphology of the system, and that also explains its position and radial velocity.

In the
mini-bullet scenario of \S\,\ref{bullet.sec}
the ionized gas in between the nuclei is shocked and compressed due to the
recent collision, and
it may be that the black hole 
formed through the runaway gravitational
collapse of a cloud or filament within this gas.
This scenario links the SMBH to the gas cloud in which it is embedded, and explains why
its radial velocity is exactly in between the velocities of the gas in the two rings.

This idea is qualitatively similar to `heavy seed' formation models that have long been proposed
for the origin of SMBHs in the centers of galaxies. While the leading model for the origin of SMBHs is that
they began as the $\sim 10^{1-3}$\,\msun\ collapsed remnants of
the first generation of Population III stars \citep[e.g.,][]{madau:01,volonteri:03},
the direct collapse of pre-galactic $\sim 10^{4-5}$\,\msun\ gas clouds is an important alternative
\citep[see][]{haehnelt:93,eisenstein:95,bromm:03,lodato:06}. 
Simulations show that the lack of metals in these early baryonic objects, combined with
external radiation fields and the violent gas dynamics associated
with the formation of the halo, can create conditions that are conducive for SMBH formation
\citep{lodato:06,natarajan:11,wise:19}.
Recently these models have received renewed attention,
as early JWST results are indicating that many
galaxies have relatively high black hole masses for their stellar mass
\citep{natarajan:17,natarajan:24,furtak:24,greene:24,matthee:24}.

The similarity to black hole seeding models
is only superficial, as the gas in the $\infty$ galaxy
is metal-rich and not at the center of the halo. Recently the first
theoretical studies of mini-bullet events have been performed,
in the context of forming dark matter-free dwarf
galaxies out of the post-collision gas \citep{silk:19,shin:20,lee:24}.
While these studies do not have the resolving power to study
SMBH formation, they
do indicate that regular star formation is suppressed in
the turbulent post-collision gas, while
the formation of massive self-gravitating clumps is promoted \citep{silk:19,lee:21}.
Furthermore, high resolution studies of the aftermath of gas-rich mergers 
have shown that black holes may form in the central regions of
the remnant, even though the gas is metal-rich \citep{mayer:10,mayer:15}.
In this formation channel turbulence and thermal pressure,
rather than the absence of metals, prevent fragmentation and star formation.
The gas in the $\infty$ galaxy is currently spread over a $\sim 10$\,kpc
region, but it is conceivable that similar extreme conditions were reached locally at the moment
of the collision of the two progenitor galaxies.

We note that, in this scenario, both nuclei still have their own, very massive, SMBHs. The stellar velocity dispersions of the nuclei are likely $\sim 300$\,\kms\ based on their $\sim 1$\,kpc sizes and masses of $\sim 10^{11}$\,\msun. The $M_{\rm BH} - \sigma$ relation implies black hole masses of $\sim 10^9$\,\msun\ for dispersions in this range \citep{mcconnell:13}. It would be interesting to look for AGN activity in the two nuclei with sensitive radio or spectroscopic observations, as that could confirm that there are three black holes in the system.

\begin{figure}[htbp]
  \begin{center}
  \includegraphics[width=0.7\linewidth]{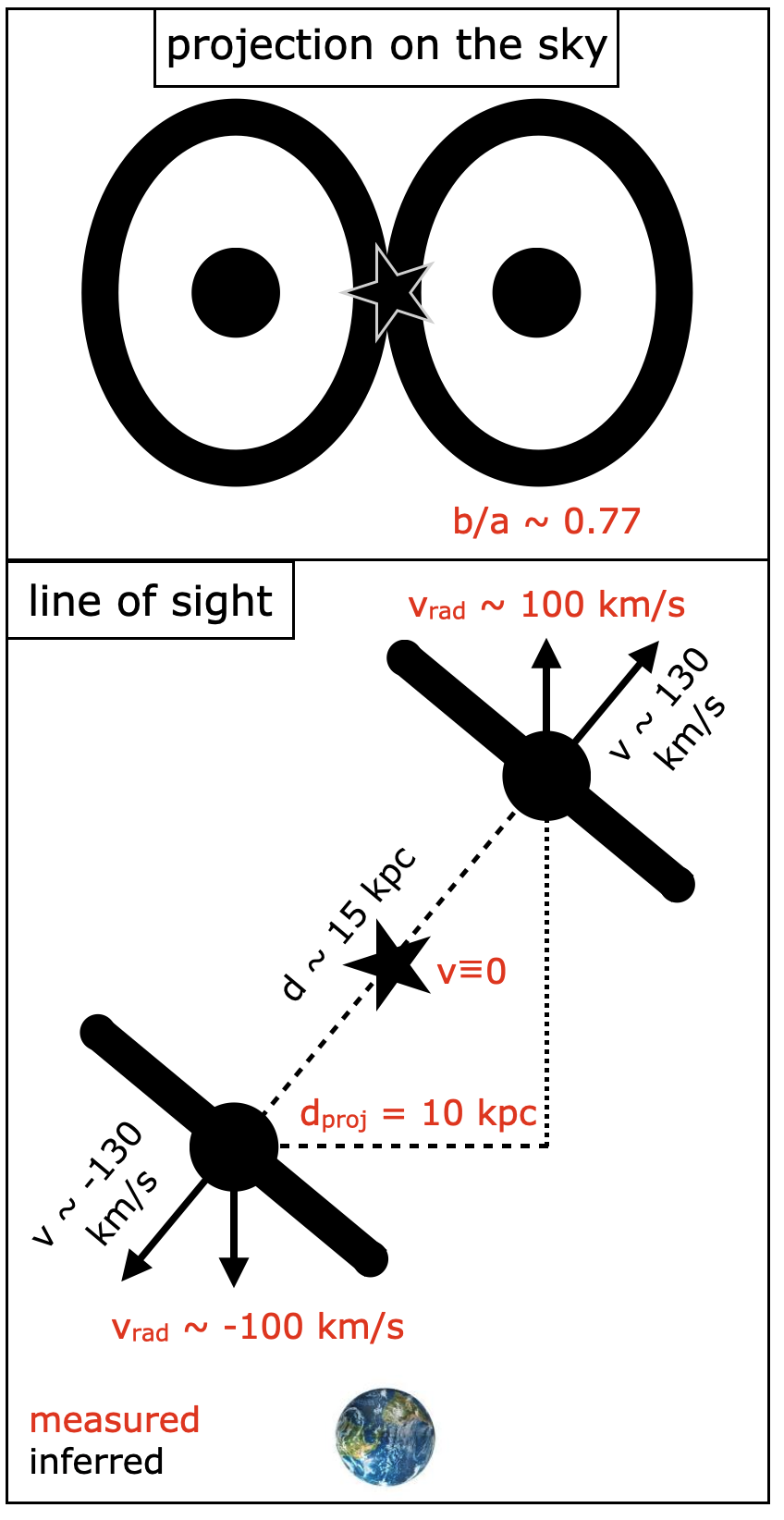}
  \end{center}
    \caption{\small 
{\em Main panel:} Possible geometry of the $\infty$ galaxy along
the line of sight. Measured properties are indicated in red and derived
properties in black. The orientation is set by the axis ratio of the
rings, $b/a \sim 0.77$.  {\em Top panel:} Projection onto the sky.
The rings appear to be overlapping but are actually $\sim 15$\,kpc apart.
   }
   \label{geometry.fig}
    \vspace{-1pt}
\end{figure}

\subsection{Geometry of the System and Initial Black Hole Mass}

As discussed in \S\,\ref{bullet.sec}, the unusual morphology
of the $\infty$ galaxy can be explained by
a face-on collision of two disk galaxies,
leading to the formation of collisional rings around the two surviving
bulges.  In this context we can use the observed properties of the
galaxy to determine its approximate 3D geometry. Assuming that the rings are
intrinsically round, their observed axis ratio of $b/a \approx 0.77$
implies an angle with respect to the plane of the sky of $\approx 40^{\circ}$.
The physical distance between the two nuclei is then $\approx 15$\,\,kpc.
The proposed geometry along the line of sight is shown in Fig.\ \ref{geometry.fig}.

We can use the deprojected geometry to estimate $\Delta t$, the time that has
elapsed since collision.
The deprojected radial velocity difference between the SE and NW sides of the
system is approximately $260$\,\kms\ (see Fig.\ \ref{geometry.fig}),
and for a deprojected separation of 15\,kpc this gives $\Delta t
\sim 50$\,Myr. 
This time interval is consistent with the typical time scales for collisional
ring formation in simulations \citep{struck:10}.
The implicit assumption here is that the measured
[O\,II] velocities in the SE and NW are a proxy of the systemic velocities
of the $\infty$SE and $\infty$NW nuclei. This is unlikely to be the case,
as the measured line-of-sight velocities of the rings 
are expected to be a combination of the post-collision velocity of the system, the outward density wave,
and the shock dynamics and angular momentum redistribution during the collision
\citep{higdon:96}.
We also assume
that the nuclei have not yet turned around and are
on their initial post-collision trajectories.

In the context of the direct-collapse model,
the elapsed time gives us a rough estimate of the initial mass
of the SMBH. For a standard radiative efficiency of $\eta \approx 0.1$,
a black hole that accretes at the Eddington rate increases its mass by a factor
of $\sim 3$ over 50\,Myr. For a current
mass of $10^6$\,\msun\ this gives an initial mass of $\sim 3\times 10^5$\,\msun.
We note that, judging from the spectral index,
the currently observed episode of black hole activity probably commenced  much more
recently (see \S\,\ref{agn.sec}). This indicates that the
accretion was probably stochastic over the past $\sim 50$\,Myr.

\section{Conclusions}

In this paper we have presented an unusual galaxy system, consisting of two nuclei with rings in a striking
symmetric configuration. In between the nuclei is a SMBH with quasar-like levels of activity
that is embedded in a distribution of ionized
gas. We suggest that the presence of the SMBH in this system and at that location is not coincidental, but
the result of a causal chain of events: 1) a nearly face-on
collision between two disk galaxies with a small impact parameter, like the one that produced II\,Hz\,4
(see \S\,\ref{binary.sec}); 2) the interaction of gas clouds at the collision site, leading
to shocks and compression in a process
akin to what is seen in the bullet cluster; 3) the runaway collapse of a dense cloud of gas
into a black hole at the collision site; and 4) accretion onto this black hole from the surrounding gas.

The proposal that black holes can form at late times in interacting galaxies
is not new; as an example, \citet{schawinski:11} 
suggested that the presence of several AGNs in a clumpy $z=1.35$ galaxy could be due to in-situ
formation and late seeding.
The active black hole in the $\infty$ galaxy stands apart in two important ways: it is perhaps the clearest example
yet of a SMBH that is outside of a galaxy nucleus, and we propose a specific mechanism for
its formation that can be tested with simulations and follow-up observations.

The mini-bullet collision can be
simulated with strong observational
constraints on the post-collision conditions (such as the positions of the nuclei, the morphology
of the rings, the location and morphology of the gas, and the observed radial velocities).
It may be that an in-depth
analysis of the physical conditions in the colliding clouds will
demonstrate that SMBHs cannot form in this scenario. In that case, we are probably witnessing the (re-)ignition
of a wandering or ejected SMBH as it passes through the gas in the central regions of the $\infty$ galaxy.
If it does turn out to be possible to form black holes, we will learn a lot about the process.
For instance, it may be that
the collapse is hierarchical, with mergers of massive stars leading to
the formation of intermediate-mass black holes and
multiple intermediate-mass black holes merging to form the black hole that we now detect
\cite[see, e.g.,][]{ebisuzaki:01}.

Other tests will come from observations.
The spectroscopy presented in this paper is limited
in its spatial resolution and does not cover the key optical emission lines H$\alpha$, [N\,II], and [S\,II].
These lines are inaccessible from the ground due to H$_2$O absorption in our atmosphere, but they can be observed with
JWST. With the NIRSPEC integral field unit, the presence of the line-emitting gas in between the nuclei could
be confirmed, the radial velocities of the nuclei could be measured directly, and the predicted transition
between photo-ionization close to the black hole to shock-ionization further out could be observed.
Furthermore, any radial velocity offset between the black hole and the surrounding gas could be accurately
measured. The most compelling evidence for a runaway gravitational
collapse of a clump within this gas would be the observation that there is {\em no}
offset: as noted in \S\,\ref{insitu.sec} this would be difficult to reconcile with a wandering or ejected
black hole, and it is a prediction of the in-situ formation model.

If our proposed scenario is
confirmed, the $\infty$ galaxy provides an empirical demonstration that direct-collapse formation
of SMBHs can happen
in the right circumstances -- something that has so far only been seen in simulations and through indirect observations
\citep[such
as high SMBH masses in high redshift galaxies;][]{greene:24,furtak:24,natarajan:24,jeon:25}.

\begin{acknowledgements}
This paper uses HST data from program 9822 and JWST data from programs 1837 and 2561, accessible using DOI
\dataset[10.17909/r5ab-6953]{https://doi.org/10.17909/r5ab-6953}.
The VLA 3\,GHz data are available using DOI \dataset[10.26131/IRSA176]{https://doi.org/10.26131/IRSA176}.
The notebooks that were used to reduce the data are available as part of the
{\tt grizli} package, on Zenodo: \dataset[10.5281/zenodo.1146904]{https://doi.org/10.5281/zenodo.1146904}.
\end{acknowledgements}

\bibliography{master_0525}{}
\bibliographystyle{aasjournal}

\begin{appendix}

\begin{figure*}[htb]
  \begin{center}
  \includegraphics[width=0.9\linewidth]{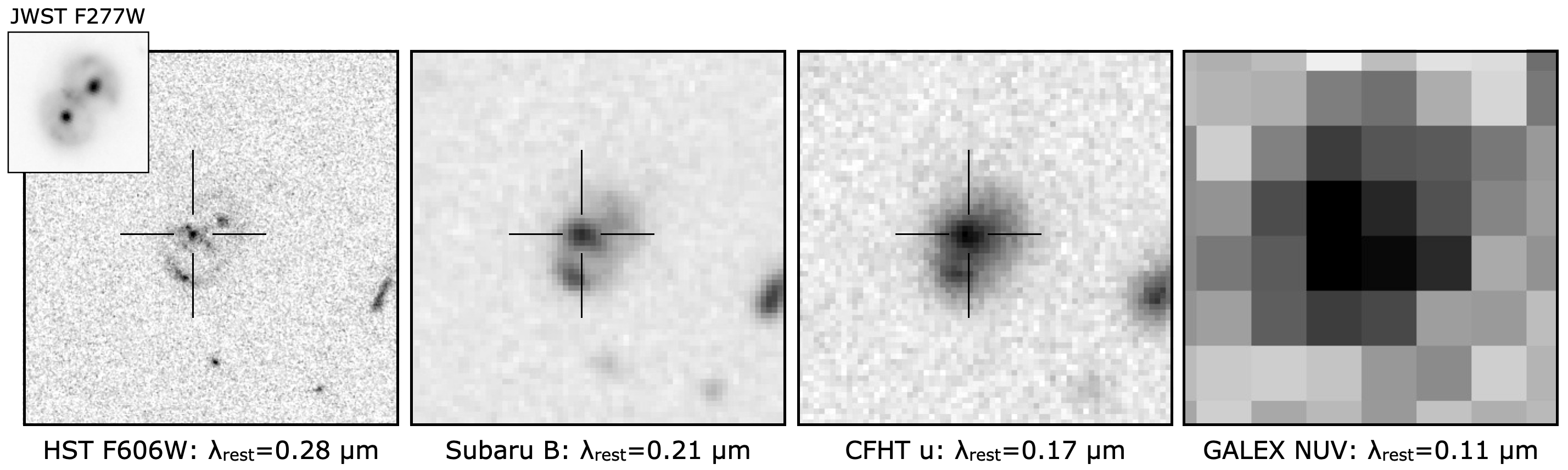}
  \end{center}
    \vspace{-0.3truecm}  
    \caption{\small The $\infty$ galaxy in the rest-frame ultraviolet. 
From left to right, images at progressively shorter wavelengths are shown. For reference, the inset
shows the JWST F277W image ($\lambda_{\rm rest} = 1.3\,\mu$m). The location of the blue
compact object that we associate with the black hole is indicated with a cross. It is bright
in the Subaru $B$ and CFHT $u$ bands, and likely dominates the GALEX NUV band.
   }
   \label{uv_images.fig}
    \vspace{-1pt}
\end{figure*}

\section{Ionizing Flux}

The black hole is in the center of the complex $\infty$\,cen region, as determined
from the VLA 3\,GHz localization. This region is
the equivalent of the narrow line region of an AGN, and we infer from the
Keck/LRIS spectrum that it is ionized by highly energic photons (see \S\,\ref{agn.sec}).
Here we investigate the source of these photons, and specifically whether
the $\infty$\,cen region has a sufficiently high UV luminosity to produce them.
First, we note that there
is  a compact object
at the location of the black hole, within the $\infty$\,cen
region. It is blue, and most clearly seen in the HST/ACS F606W image of the
galaxy (see Fig.\ \ref{jwst_im.fig}). We suggest that this blue object represents
radiation from the accretion disk of the black hole.
There are no high resolution images of the galaxy
at shorter wavelengths, but as
we show in Fig.\ \ref{uv_images.fig} the $\infty$\,cen region
dominates the flux of the entire galaxy in the observed $B$ and $u$ bands, at
rest-frame wavelengths of $0.21\,\mu$m and $0.17\,\mu$m respectively.
Remarkably, the galaxy is clearly detected in the GALEX NUV band at $\lambda_{\rm rest}\approx 0.11\,\mu$m,
close to the ionization edge of 912\,\AA. The GALEX image is unresolved, but
based on the morphology in the $B$ and $u$ bands it is
dominated by the $\infty$\,cen region, and plausibly by
the compact object within that region.

In the following, we use the observed GALEX NUV flux density to calculate the rate of ionizing photons
in the $\infty$\,cen region. Some of the rest-frame far-UV light likely comes from other regions, such as 
H\,II regions in the rings or the shock front between the nuclei,
which may lead us to overestimate the
ionizing flux that is coming from within
the $\infty$\,cen region. However, the larger effect is dust, which
works in the opposite direction: the intrinsic far-UV luminosity is almost certainly
significantly higher than the observed, dust-attenuated
luminosity. Given the visible dust lanes in the image (see Fig.\ \ref{jwst_im.fig}), and
the $A_V\sim 2$ attenuation toward the two nuclei, the detected GALEX flux is probably
attenuated by a factor of $10-20$.

The GALEX NUV AB magnitude $m_{\rm NUV} = 24.7\pm 0.1$, corresponding to a rest-frame luminosity
density of $L_{\nu} \approx 1.5 \times 10^{-13}$\,erg\,s$^{-1}$\,Hz$^{-1}$
at $\approx 1100$\,\AA. The ionization of [O\,III]
requires photons with energy $>35.1$\,eV, corresponding
to wavelengths $<352$\,\AA. The ionizing photon production rate is therefore given by
\begin{equation}
Q = \int_{35.1\,{\rm eV}}^{\infty} \frac{L_{\nu}}{h\nu} d\nu.
\end{equation}
Assuming
a powerlaw in the UV with $L_{\nu} \propto \nu^{-1.5}$,
this gives $Q\sim 5\times 10^{53}$\,photons\,s$^{-1}$.
The [O\,III] luminosity $L_{\rm [O\,III]} = f h\nu_{\rm [O\,III]} Q$, with $f$ the fraction
of ionizing photons that contribute to [O\,III]. Comparing this
to the observed [O\,III] luminosity of $\approx 1.2\times 10^{42}$\,erg\,s$^{-1}$, we
find that the GALEX-detected UV source can produce the LRIS-measured line luminosity
for $f\sim 0.1$. If the UV source is attenuated by a factor of $\sim 10$, then $f\sim 0.01$.

\section{The Binary Collisional Ring Galaxy II\,Hz\,4}
\label{binary.sec}

\begin{figure*}[t]
  \begin{center}
  \includegraphics[width=0.65\linewidth]{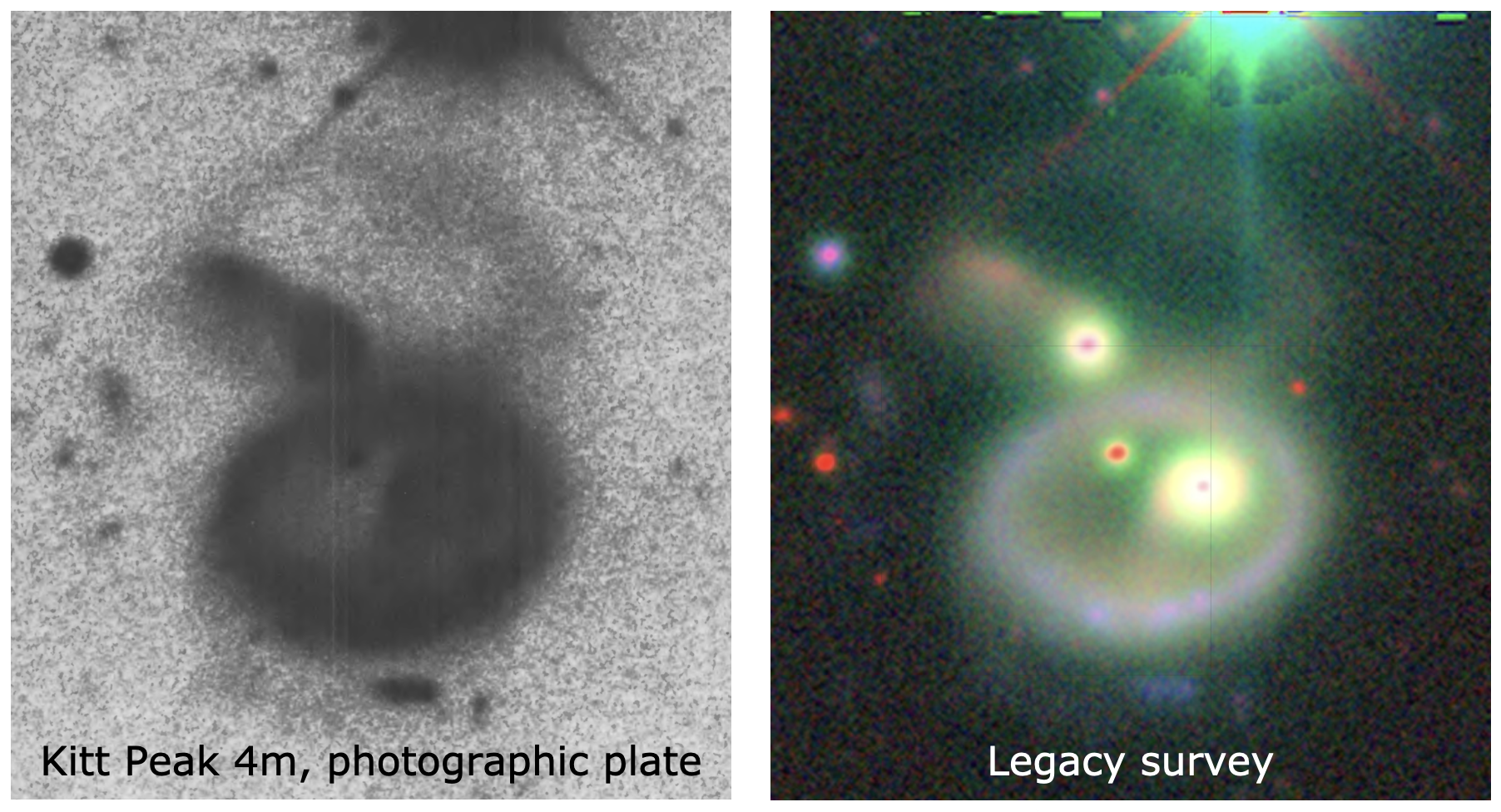}
  \end{center}
    \vspace{-0.3truecm}  
    \caption{\small 
{\em Left:} Deep photographic plate of the
binary ring galaxy II\,Hz\,4,  obtained with the Kitt Peak 4m
and reproduced from \citet{lynds:76}.
{\em Right:} Legacy survey image of the galaxy. 
   }
   \label{IIHz4.fig}
\end{figure*}

The closest known analog to the $\infty$ galaxy is II\,Hz\,4, a galaxy with two rings and
two bulges at $z=0.043$. 
We reproduce the deep photographic plate of II\,Hz\,4 obtained
by \citet{lynds:76} in Fig.\ \ref{IIHz4.fig},
along with the $grz$ Legacy survey image of the galaxy \citep{dey:19}.
The entire system is larger than the $\infty$ galaxy,
with a total spatial extent along the long axis of $\sim 40$\,kpc, but is otherwise quite
similar. The two bulges and a portion of
the bright ring have SDSS spectroscopy \citep{sdss1}. The bulges have early-type
spectra with $z=0.04288$ (South) and $z=0.04295$ (North) respectively. The ring has strong emission
lines and has $z=0.04296$. These values are consistent with the original
measurements of \citet{lynds:76}, who also note that the Northern ring does not have
strong emission lines. It would be interesting to obtain further observations of this object.

\begin{figure*}[t]
  \begin{center}
  \includegraphics[width=0.95\linewidth]{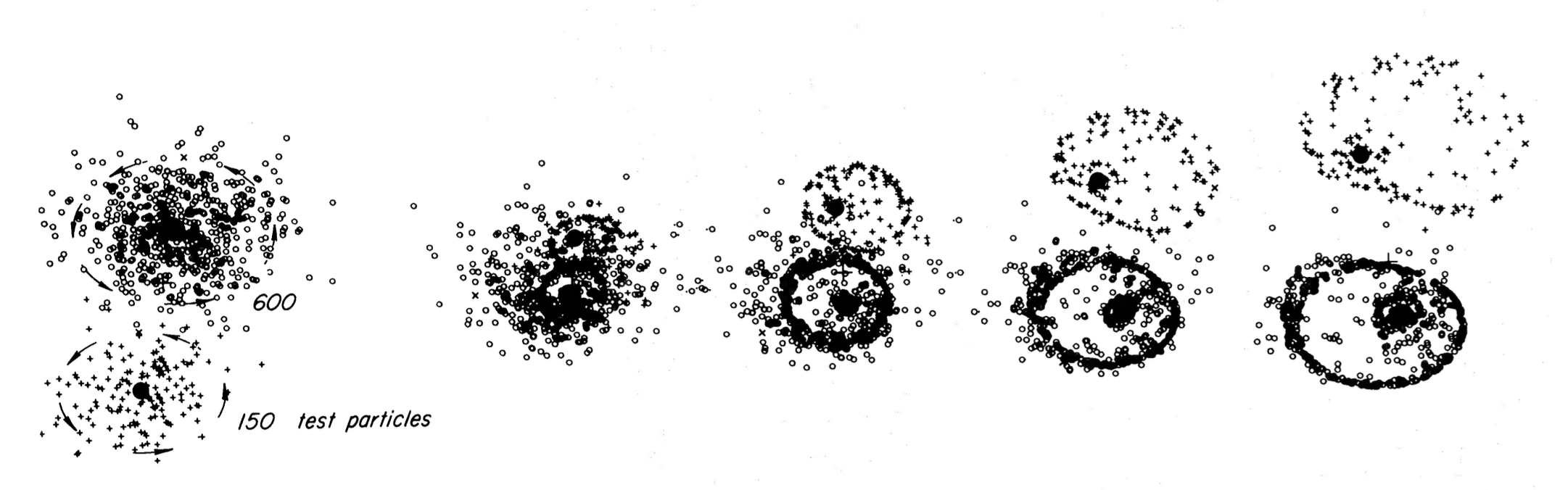}
  \end{center}
    \vspace{-0.3truecm}  
    \caption{\small 
N-body simulation of II\,Hz\,4, reproduced from \citet{lynds:76}. In a collision of two bulge\,+\,disk
galaxies with a small impact parameter disk stars are herded in outwardly expanding rings around
both bulges.
   }
   \label{lynds_fig6.fig}
\end{figure*}

II\,Hz\,4 was modeled as a mutual collisional system by
\citet{lynds:76}, in the first successful N-body simulation done for any ring galaxy.\footnote{In fact, its
binary ring morphology led the authors to the now-standard explanation that collisional rings are
caused by a compact impactor.}
Lynds \& Toomre demonstrated that collisions of bulge\,+\,disk systems can lead
to the herding of disk stars into outwardly expanding rings around both bulges. In the
specific simulation shown in Fig.\ \ref{lynds_fig6.fig}
the rotation axes of the disks are parallel to the direction of motion of
the galaxies, although the authors note that rings can also be produced
if the disks are misaligned by up to 30$^{\circ}$.
Of greater importance is the impact parameter; to produce a symmetric response the bulges
must pass near each other and interior to most of the disk stars.

\end{appendix}

\end{document}